\documentclass[12pt]{article}
\usepackage{amsmath}%
\usepackage{amsfonts}%
\usepackage{amssymb}%
\usepackage{graphicx}

\begin{document}

\title{A 2-Dimensional Cellular Automaton for Agents Moving from
Origins to Destinations}
\author{Najem Moussa \thanks{e-mail: najemmoussa@yahoo.fr}
\\\textit{EMSN, D\'{e}pt. de Physique, FST, }
\\\textit{B.P. 509, Boutalamine, Errachidia, Morocco}}
 \maketitle

\begin{abstract}
We develop a two-dimensional cellular automaton (CA) as a simple
model for agents moving from origins to destinations. Each agent
moves towards an empty neighbor site corresponding to the minimal
distance to its destination. The stochasticity or noise ($p$) is
introduced in the model dynamics, through the uncertainty in
estimating the distance from the destination. The friction
parameter $"\mu"$ is also introduced to control the probability
that the movement of all agents involved to the same site
(conflict) is denied at one time step. This model displays two
states; namely the freely moving and the jamming state. If $\mu$
is large and $p$ is low, the system is in the jamming state even
if the density is low. However, if $\mu$ is large and $p$ is high,
a freely moving state takes place whenever the density is low. The
cluster size and the travel time distributions in the two states
are studied in detail. We find that only very small clusters are
present in the freely moving state while the jamming state
displays a bimodal distribution. At low densities, agents can take
a very long time to reach their destinations if $\mu$ is large and
$p$ is low (jamming state); but long travel times are suppressed
if $p$ becomes large (freely moving state).
\newline\ Pacs numbers: 45.70.Vn, 02.50.Ey, 05.40.-a
\newline\ \textit{Keywords:} Traffic;
freely moving; jamming; clustering; travel times.
\end{abstract}

\section{Introduction}

Cellular Automata (CA) micro-simulation has emerged as a tool for
simulating traffic flow and modelling transport networks
\cite{Hel1,chow,nag}. In CA, time and space are discrete. The
space is represented as a uniform lattice of cells with finite
number of states, subject to a uniform set of rules, which drives
the behavior of the system. These rules compute the state of a
particular cell as a function of its previous state and the state
of the neighboring cells.
\newline\ Agents moving from origins to destinations across networks
may represent several real entities, as for example: ants,
biological organisms, small robots, transport in micro-mechanical
systems, crowd flow, packets transport in the Internet, etc.... It
was found that the motion of the biological organisms is usually
controlled by interactions with other organisms in their
neighborhood and randomness also plays an important role
\cite{Vicsek}. Real ants have been shown to be able to find
shortest paths towards destinations using as only information the
pheromone trail deposited by other ants \cite{Beck}. The finding
of shortest paths, inspired from ants' behavior, are successfully
applied to several physical problems such as, pedestrians
\cite{Bur, Kir}, traffic flow \cite{Mitch}, combinatorial
optimization \cite{Dorigo} and circuit switched communications
network problems \cite{Schoon}. The problems of movement of agents
with origins and destinations were studied using a two dimensional
cellular automata \cite{Mitch,Man}. In these models an agent tries
to reach its destination using simple rules. Transitions from the
freely moving to the jamming states were studied. A variant
2-dimensional CA model for simulation of agents moving from
origins to destinations will be presented here. Agents moving
across the network have sensors to perceive their local
neighborhood and their destinations and then affect their
environment. This is done especially by estimating the distance
metric to the destination site. The concern herein will be with
the movement, propagation, and interaction of agents in low and
high density situations. This will be done by exploring the
patterns and behaviors of the spatio-temporal organization of
agents. The objective of this research is to provide insight into
modelling complex dynamics using CA microsimulation and capturing
general features of agents travelling from origins to
destinations. The paper is organized as follows. In Sec. $2$, we
describe our model for movements of agents with origins and
destinations. In Sec. $3$, we present our numerical results where
we give the phase diagrams of the system. A detailed description
of the cluster size and the travel time distributions are also
presented. Finally, we conclude with some conclusions in Sec. $4$.

\section{The cellular automata model}
The CA model is a two-dimensional cellular automaton based on a
square lattice with periodic boundary conditions. There is a fixed
number of agents on the lattice at all times. Only one agent at
most can occupy a given site. At any time step, an agent can move
at most to one of its 4 neighboring sites. Updating of the CA
occurs in parallel where the rules are applied to all agents at
the same time. Agents are associated with given origin-destination
sites. The origin and destination sites must be different. An
agent travels from the origin site towards its destination site,
whereupon it disappears. Each disappeared agent is immediately
replaced by a new agent, and so the agent number is always
constant in the lattice. A new origin-destination pair is then
chosen randomly for this new agent. If, however, there is an agent
already present at the chosen origin, then another origin site is
selected.
\newline\ Agents will move towards their destinations at all
times by selecting an unoccupied neighboring site which has the
minimal distance from that site to the destination site (see
figure 1). An agent examines the unoccupied neighboring sites. For
each of these sites a distance to the destination is evaluated.
Then, a site with the minimal distance is selected as the next
site to which the agent will move. If all neighboring sites are
occupied it will not move.
\newline\ The stochasticity or noise is introduced in the model
dynamics, through the uncertainty in estimating the distance from
the destination. So, with probability $p$ an agent moves towards
an arbitrary empty neighboring site rather than the site of
minimal distance. The friction parameter $\mu$ is also introduced
to control the probability that the movement of all agents
involved to the same site (conflict) is denied at one time step.
This friction parameter which is essential for resolving the
conflict arising in parallel update simulations, is applied for
pedestrian traffic problems \cite{Kir2,Kir3}.
\newline\ In each time step, positions, speeds and directions of
all agents are updated according to the following local rules:
\newline\ - with probability $p$, an agent selects one arbitrary
empty neighboring site.
\newline\ - with probability $(1-p)$ agent
selects an empty neighboring site corresponding to the minimal
distance to the destination. If two empty neighboring sites of one
agent have the same minimal distance from the destination then one
of these two allowed neighbors is chosen randomly.
\newline\ If two or more agents select the same
site (conflicts) then:
\newline\ - with probability $\mu$ none of the agents is allowed
to move to their selected site.
\newline\ - with probability $(1-\mu)$ one of these agents is chosen
randomly to move to its selected site; the others agents do not
move.
\newline\ If there exist no conflict, the agent moves to its selected
site. If all neighboring sites are occupied, the agent does not
move.

\section{Simulation experiments and results}
We carry out our computer simulations of the model by considering
a square lattice of size $L$ with periodic boundary conditions.
Initially, we put randomly a number $N$ of agents into  the
lattice. The density of agents is denoted as $\rho=N/L^{2}$. The
velocity of each agent can be either 1 or 0. The duration of each
simulation run is $50,000$ time steps with the first $20,000$ time
steps to initiate the simulation and the latter $30,000$ used to
generate performance statistics. Agents are only allowed to move
to unoccupied nearest neighbor sites in one time step, i.e.
$v_{max}=1$ cell/time step.
\subsection{Diagrams of agents speed versus density}
In figure 2, we carried out the plots of the mean velocity of
agents as a function of the density, for several system size.
Hence, the plots show that $\langle v \rangle$ undergoes a sudden
phase transition from a freely moving state $\langle v
\rangle\approx1$ to jammed state $\langle v \rangle\approx0$ at a
critical density $\rho_{c}$. In the freely moving state,
interaction between agents is weak and the propagation is
important inside the network. In contrast, for large density, the
interaction becomes strong and jamming takes place where agent
movements become rare. As regards the variation with system size
$L$, we find that the critical density decreases with increasing
$L$ and the phase transition becomes sharper.
\newline\ In figure 3, we plot $\langle v
\rangle$ as a function of $\mu$ for different values of $p$ and
$\rho$. Hence, for low density, the average speed remains almost
constant if the probability $p$ is high. However, if $p$ is low,
$\langle v \rangle$ undergoes a sudden decrease when $\mu$ exceeds
a critical value $\mu_{c}$. This corresponds to a phase transition
from the freely moving state to the jamming state. Consequently,
the enhancement of the friction parameter can topple over from the
freely moving to the jamming of agents even at low densities. At
high densities, $\langle v \rangle$ decreases gradually with $\mu$
for all values of $p$. Yet, the speed $\langle v \rangle$ remains
always greater for larger $p$.
\newline\ The phase diagrams of the system is depicted in figure 4, where
we plotted the critical values $\mu_{c}$ as a function of $p$ for
several fixed values of $\rho$. Thus, for low densities and for a
given value of $p$, freely moving phase should exist if $\mu <
\mu_{c}$ while jamming phase takes place if $\mu > \mu_{c}$. It
was shown also from figure 4 that the jamming region is broaden as
soon as the density is increased. Yet, when the density exceeds
some value $\rho > 0.6$, the freely moving phase should never
exist.

\subsection{Spatio-temporal organizations of agents}
It is clear that the density dependence of speed alone, cannot
give the whole information on the phase behavior of the system. To
get more information on the microscopic structure of the phases,
one can determine the spatio-temporal organization of agents in
the lattice. This microscopic investigations can be obtained by
plotting the organization patterns of agents and the distributions
of cluster sizes and travel times. The cluster and the cluster
size mean here a connected bonds of unoccupied cells and a
maximally number connected cells of agents respectively. The
travel time is the time it would take to travel from the origin to
destination.
\subsubsection{Self-organization patterns of agents}
Figures 5(a-b) show a typical configurations of the organization
of agents at low density. So, for vanished value of $p$ and for
low values of $\mu$, the steady state corresponds to the freely
moving. However if $\mu$ is high, agents self-organize in a large
cluster with few freely moving agents at the boundary. For higher
values of $p$, the freely moving phase should exist even for
larger $\mu$. Hence, it seems that the role of $\mu$ is to pile up
the agents into one large cluster while $p$ tends to dispatch them
in all directions.
\newline\ For high densities and low $p$, agents pile up into one large
cluster even if $\mu$ is vanished (Fig. 5c). In the other side,
this agglomeration splits up into several clusters when $p$
becomes large (Fig. 5d).
\subsubsection{Cluster size distributions}
The cluster size distributions of the model are given in figures
6(a-b). At low density and for vanished value of $p$, only small
clusters are present in the lattice whenever $\mu$ is low. This is
one characteristic of the freely moving phase. From results
depicted in figure 6a, we observe the bimodal nature of the
cluster size distribution as $\mu$ increases. Large clusters
appear in the lattice but there are by far many more small-sized
clusters than larger ones. Furthermore, with increasing $\mu$, the
probability of small clusters diminished while that of large
cluster increases. Another important result is the discontinuity
observed from the probability distribution when $\mu$ becomes very
large. As it was shown from figure 5b, almost all agents are
congested in one large cluster with the exception of a few agents
which are located at the boundary and moving towards their
destinations. From figure 6a ($\rho=0.1$ and $p=0$), one can see a
phase transition from the freely moving phase to the jamming
phase, occurring at $\mu_{c} \approx 0.8$. Indeed, when
$\mu<\mu_{c}$, the cluster size distribution is a continuous
function; but it becomes discontinuous when $\mu>\mu_{c}$. This
value of $\mu_{c}$ agree with that found in the phase diagram
(Fig. 4). In the other hand, when $p$ is large enough, we see that
large clusters cannot exist even for large $\mu$. It is important
to note also that the increase of $p$ increases the probability of
small-sized clusters (see figure 6b, higher part).
\newline\ At high densities, it is not surprising that large clusters
may be usually present in the lattice. However, their sizes depend
greatly to the parameter $p$. For example, the most probable
cluster size (of large clusters) is shifted towards the low size
region (Fig. 6b, lower part). In contrast to $p$, the effect of
$\mu$ on the cluster size distribution is negligible (result not
shown). If the density is high, only a small value of $\mu$ may
provoke a strong congestion of agents.
\subsubsection{Travel time distributions}
The second quantity we look at is the travel time, i.e. the time
an agent needs to travel from its origin to its destination. The
probability distribution of the travel time presents a maximum
which is considered as the most probable time, at which an agent
finished its travel. At low density and for low $p$, the
distribution is sharply peaked around its maximum, whenever $\mu$
is zero (Fig. 7a). If $p$ increases, one finds evidently a broad
distribution of travel time, because agents are moving for quite
long times before reaching their destinations. The higher is $p$,
the higher is the travel time.
\newline\ As regards the variation of $\mu$, figure 7b illustrates
some graphs of the travel time distributions for low densities and
when we set $p=0$. The interesting observed phenomena is the
existence of a double asymptotic behavior. So, when $\mu$ is not
large enough, we see that infinite travel time cannot exist. In
contrast, for large $\mu$, some agent may take an infinite time to
reach its destination. In this case, the agents situated in the
interior of the large cluster do not move and rest inside for an
indefinite time. When $p$ is large enough and the density is low,
the travel time distribution changes slightly if $\mu$ is changed
(Fig. 7b). Furthermore, we see that infinite travel time does not
occur even for very large $\mu$. This shows again that the state
is the freely moving one at low density and for large $p$.
\newline\ For higher densities, the short travel times still
remain, showing the presence of some agents situated at the
boundary of a large cluster but the asymptotic behavior is rather
increasingly wide; reflecting the dynamics of agents inside the
large clusters (see figure 7a, lower part). Indeed, in the
presence of large clusters one has to distinguish between inner
and outer regions of the cluster. Inside, one finds evidently a
broad distribution of travel times, because agents are blocked for
quite long times. In the outer region of a cluster, however, one
finds shorter travel times. Furthermore, we find that infinite
travel time exist even for large $p$, because the capacity of the
freely moving is reduced when there was a big crowd of agents
(congestion).

\section{Conclusions}
In summary, we have tried to identify the behavioral aspects of
agents travelling from origins to destinations. The microscopic CA
model presented is capable of capturing the self-organization and
complex dynamics of agents. The model contains two parameters
($\mu$,$p$) and displays two states; namely the freely moving and
the jamming state. The agents speed $\langle v \rangle$ depends
strongly to the parameters ($\mu$,$p$). For low densities, phase
transitions occur as the friction parameter $\mu$ exceeds a
critical value $\mu_{c}$, which depends on both the density $\rho$
and the noise $p$.
\newline\ In the other hand, it was found that the effect of $\mu$
is to gather different agents into a large cluster. This leads to
jamming even at low density. However, the effect of $p$ is to
disperse agents through the lattice. Thus, more mobility and
fluidity will affect the whole system. When distinguished for
different density ranges, cluster size and travel time
distributions have interesting properties. Indeed, if $\rho$ and
$p$ are low, a transition from the freely moving to the jamming
states can occur at a critical value $\mu_{c}$. Thus, when
$\mu<\mu_{c}$ agents self-organized in small clusters and only
short travel times can be taken by agents. However, in the other
side ($\mu>\mu_{c}$), agents self-organize in one large cluster
with very small number of moving agents at the boundary. This is
the jamming state where the travel time can be either short and
long. So, short travel times concern agents at the boundary of the
large cluster while the long times concern those in the inner of
the cluster.
\newline\ At high densities, jamming occurs for all values of
($\mu$,$p$). This is due to the reduction of the capacity of a
freely moving when there was a big crowd of agents in the lattice.
As a result, the speed show a drastic decrease with the density.
The cluster size distribution becomes a bimodal distribution which
represents a coexistence of large clusters and small ones. Yet,
the travel time distribution is much broad where infinite travel
time exists for all values of the system parameters ($\mu$,$p$).
\newpage\

\newpage\ \textbf{Figures captions}

\begin{quote}
\textbf{Fig.1}. Illustration of agent movements in a square
lattice with periodic boundary conditions. Circles with arrows
represent agents while those without arrows represent their
destination sites. Each arrow indicates the selected site that
agent will choose (here, we set $p=0$). The conflict situation is
occurred for the "white" and the "gray" agents since they select
the same site.
\newline\ \textbf{Fig.2}. Diagrams of agents speed versus
density for different values of lattice size $L$.
\newline\ \textbf{Fig.3}. Diagrams of agents speed versus
$\mu$ for different values of $\rho$ and $p$, ($L=60$).
\newline\ \textbf{Fig.4}. Phase diagrams of the system
for several fixed values of the density: Squares, open circles,
triangles and solid circles represent $\rho = 0.1$, $\rho = 0.2$,
$\rho = 0.3$ and $\rho = 0.6$ respectively, ($L=60$).
\newline\ \textbf{Fig.5}. Self-organization patterns of
agents. Black squares represent agents. a) $\mu=0.2$, $p=0.0$ and
$\rho=0.1$, b) $\mu=0.9$, $p=0.0$ and $\rho=0.1$, c) $\mu=0.0$,
$p=0.2$ and $\rho=0.5$, d) $\mu=0.0$, $p=0.8$ and $\rho=0.5$.
\newline\ \textbf{Fig.6}. Cluster size distributions for several
values of the system parameters, ($L=60$).
\newline\ \textbf{Fig.7}. Travel time distributions for several
values of the system parameters, ($L=60$).
\end{quote}

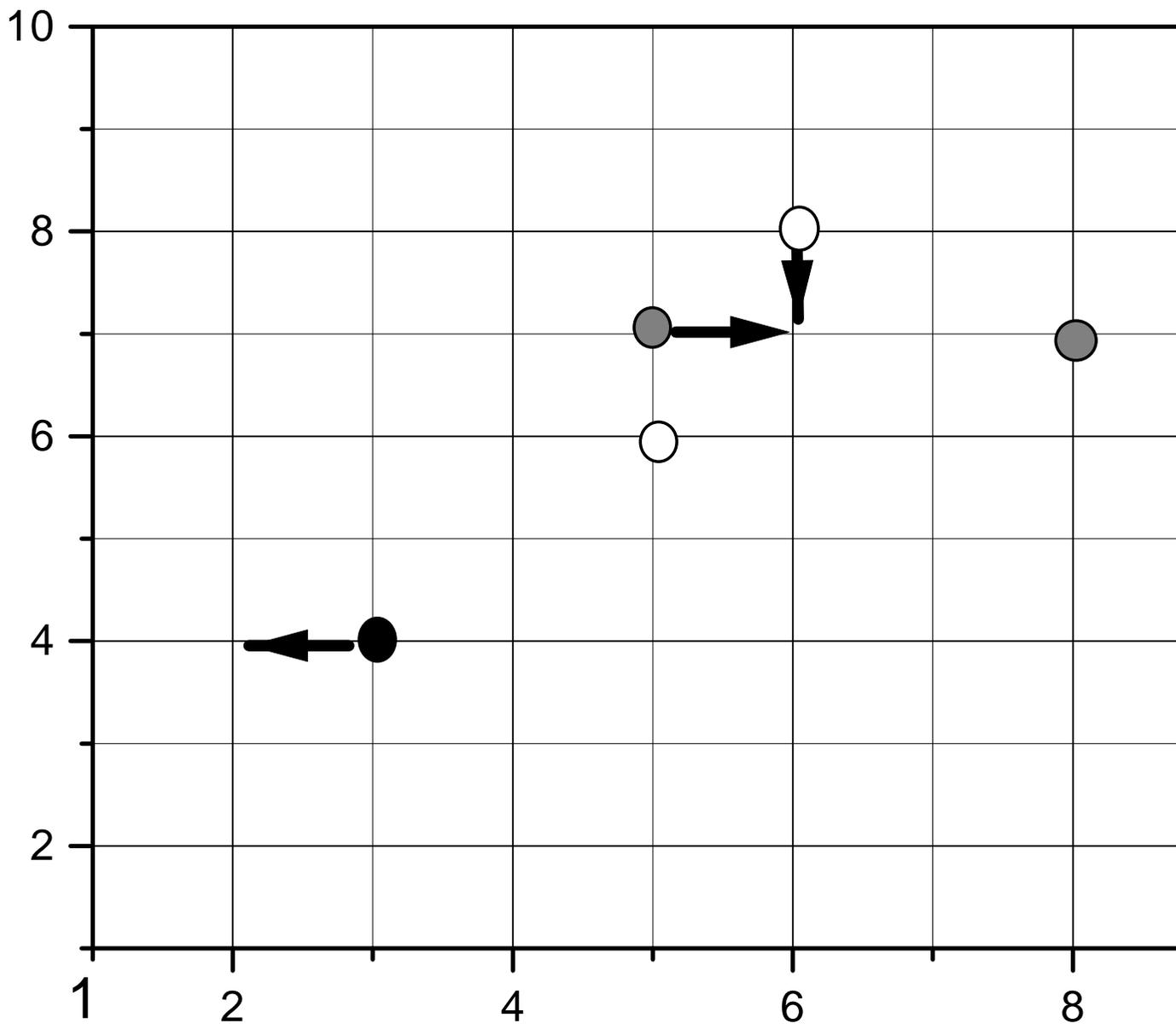
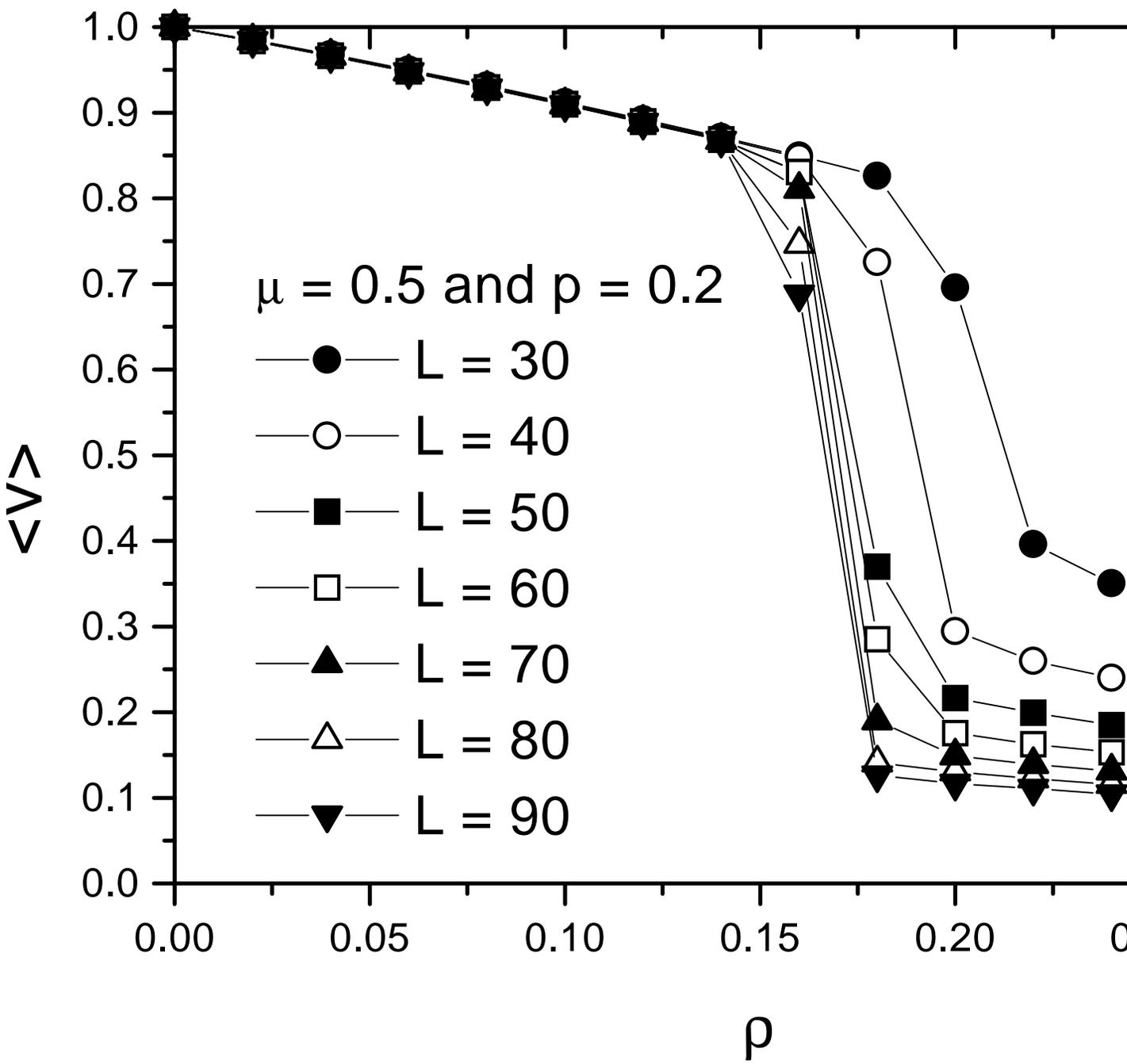
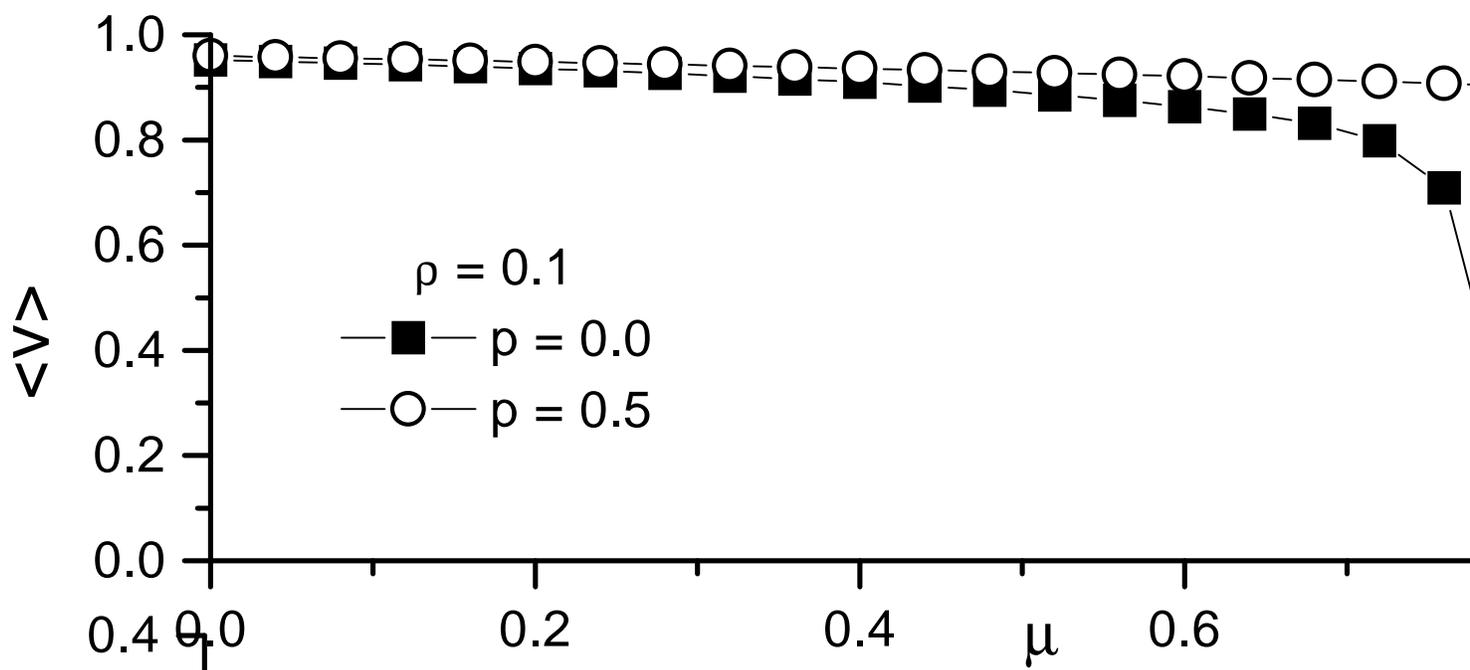
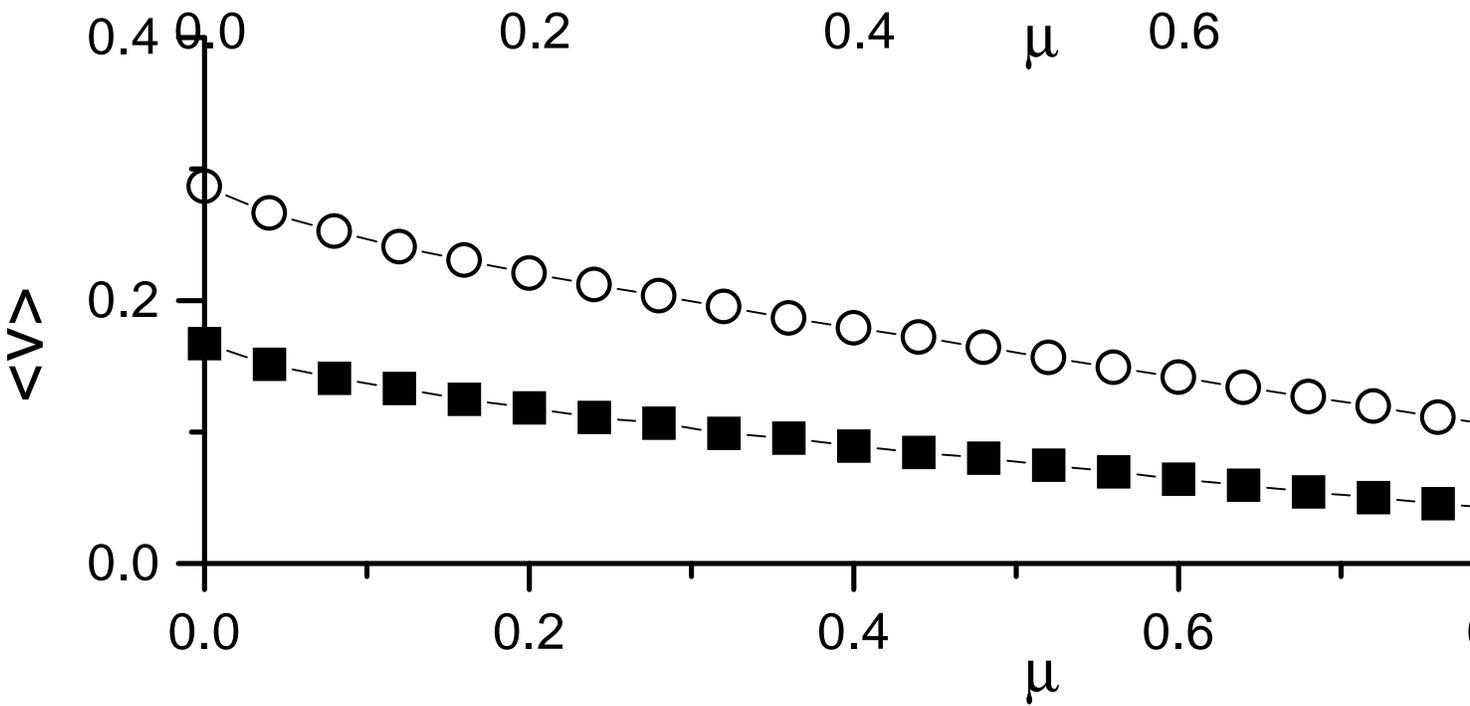
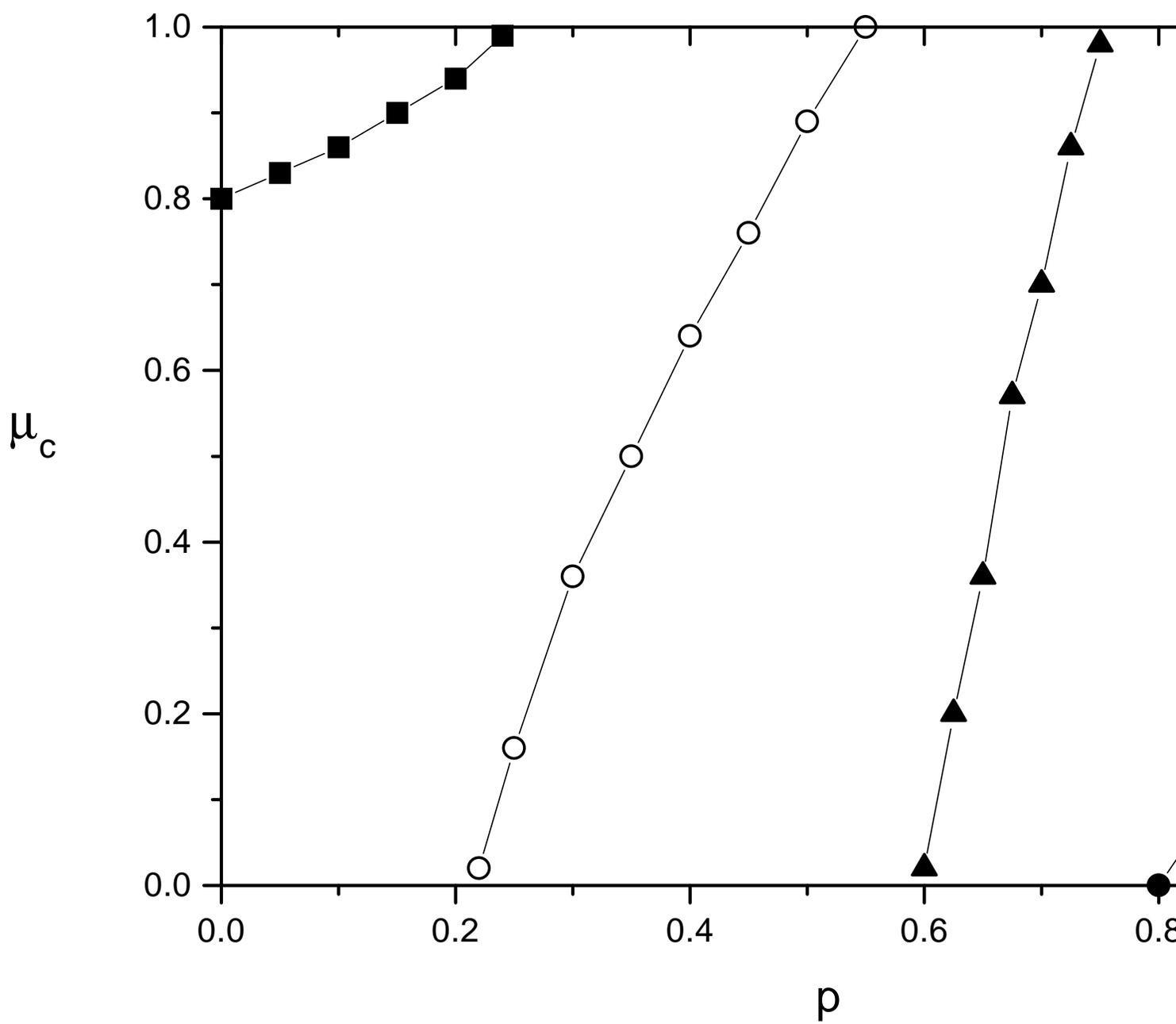
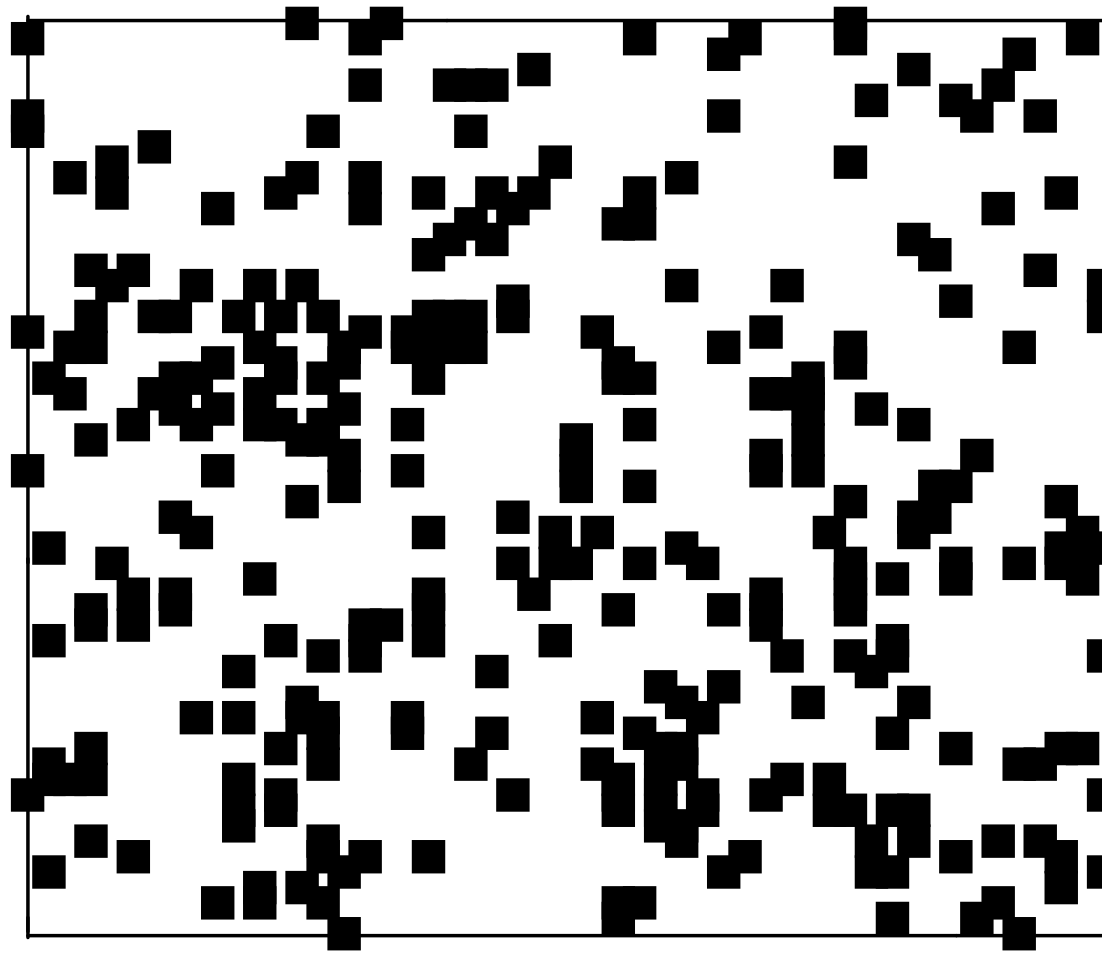
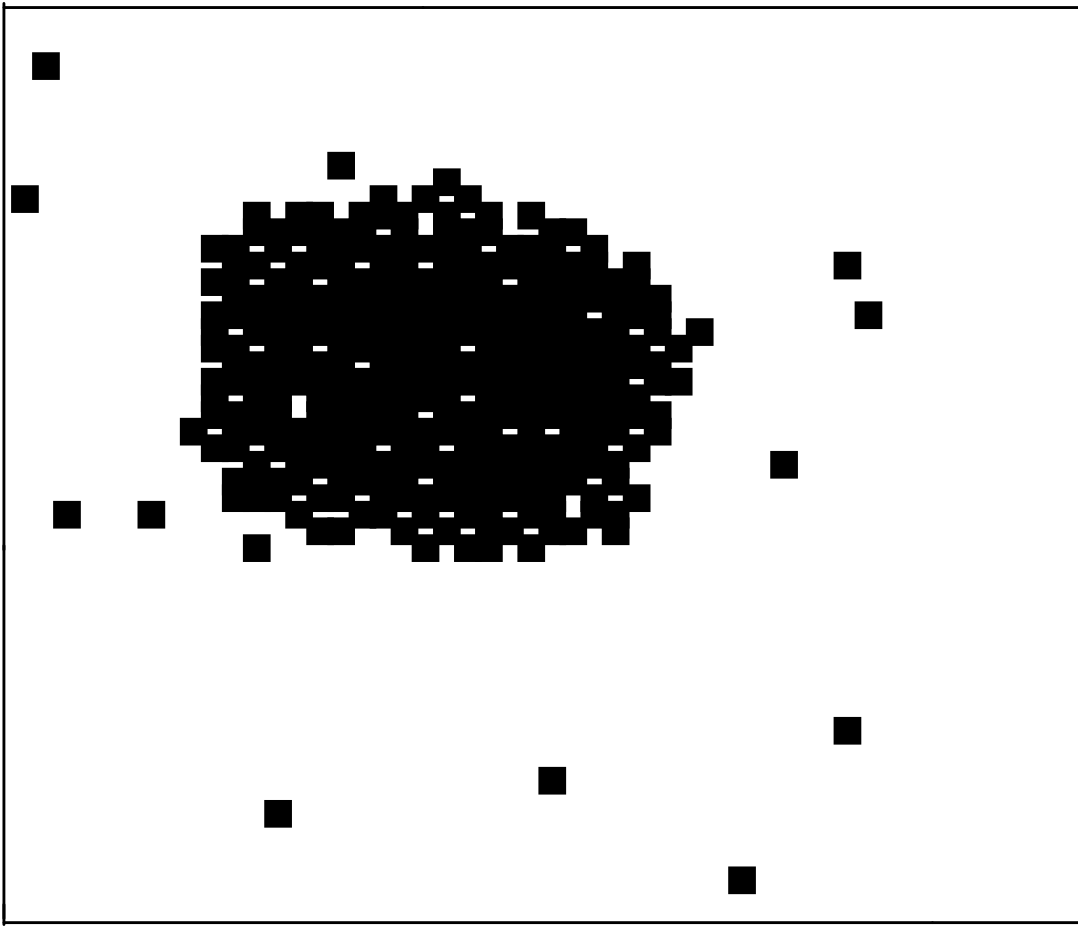
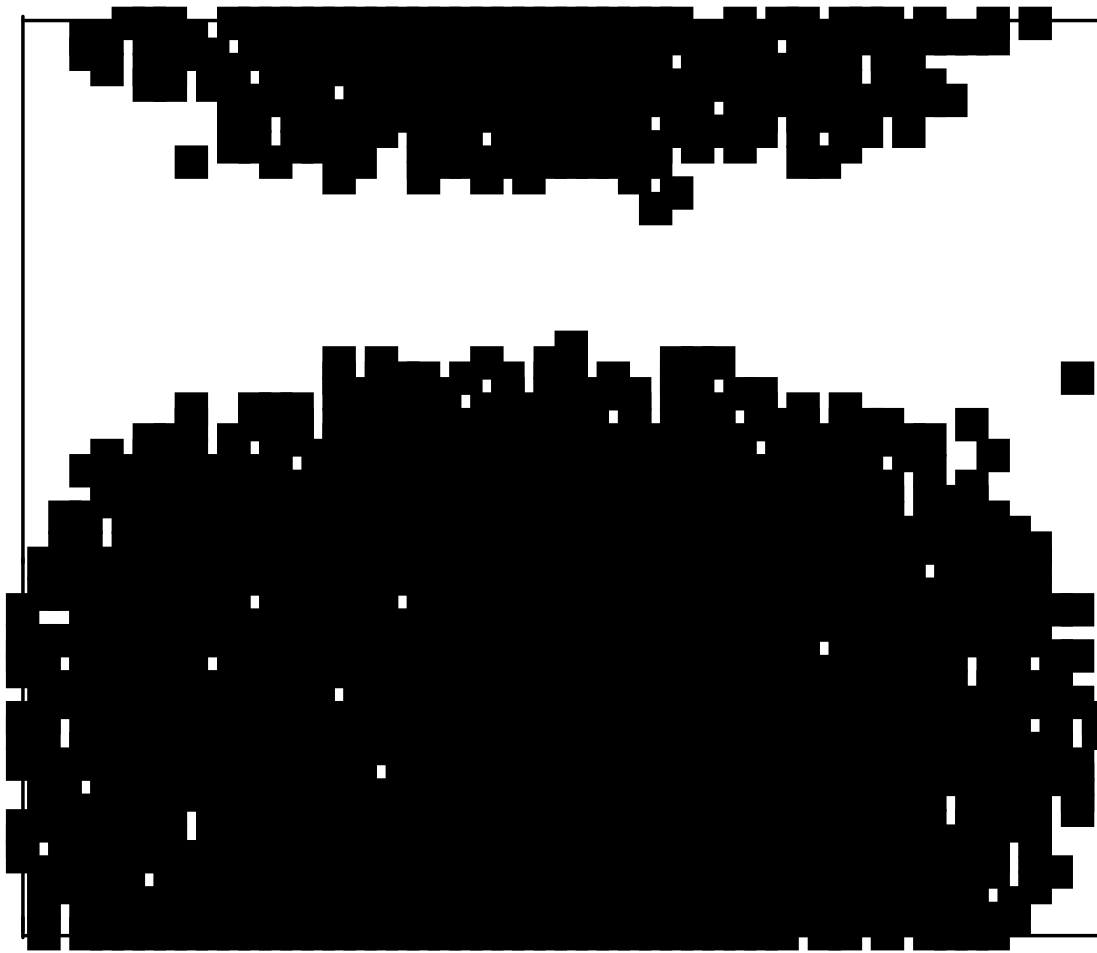
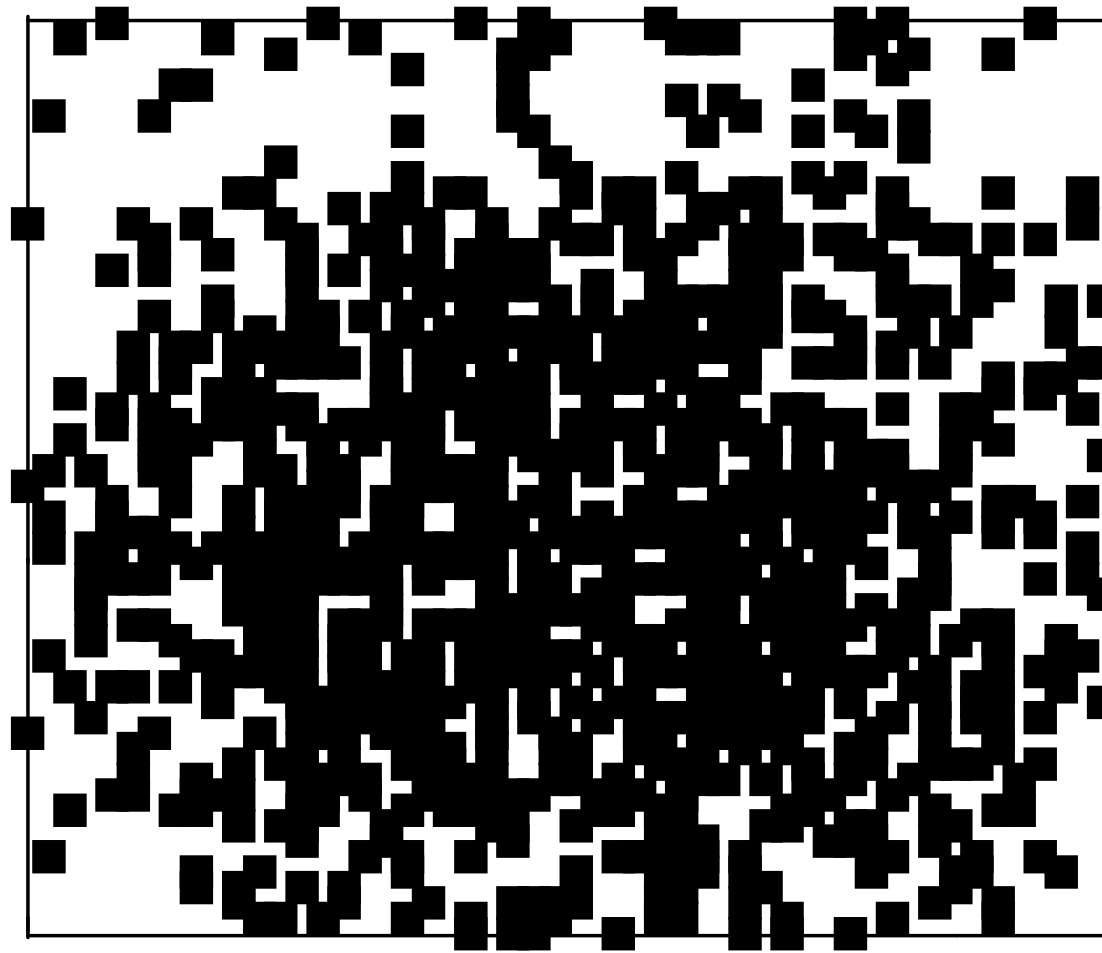
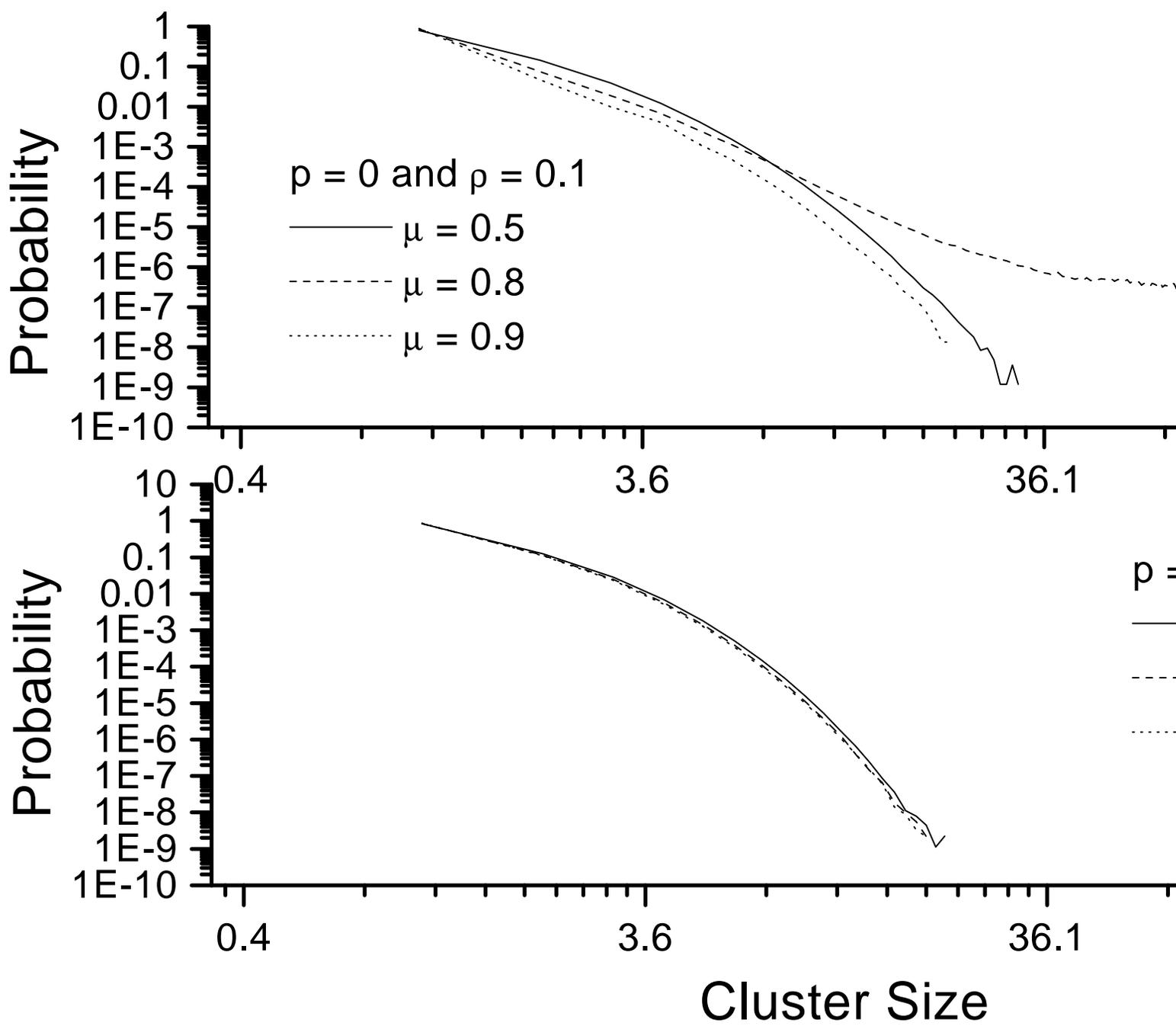
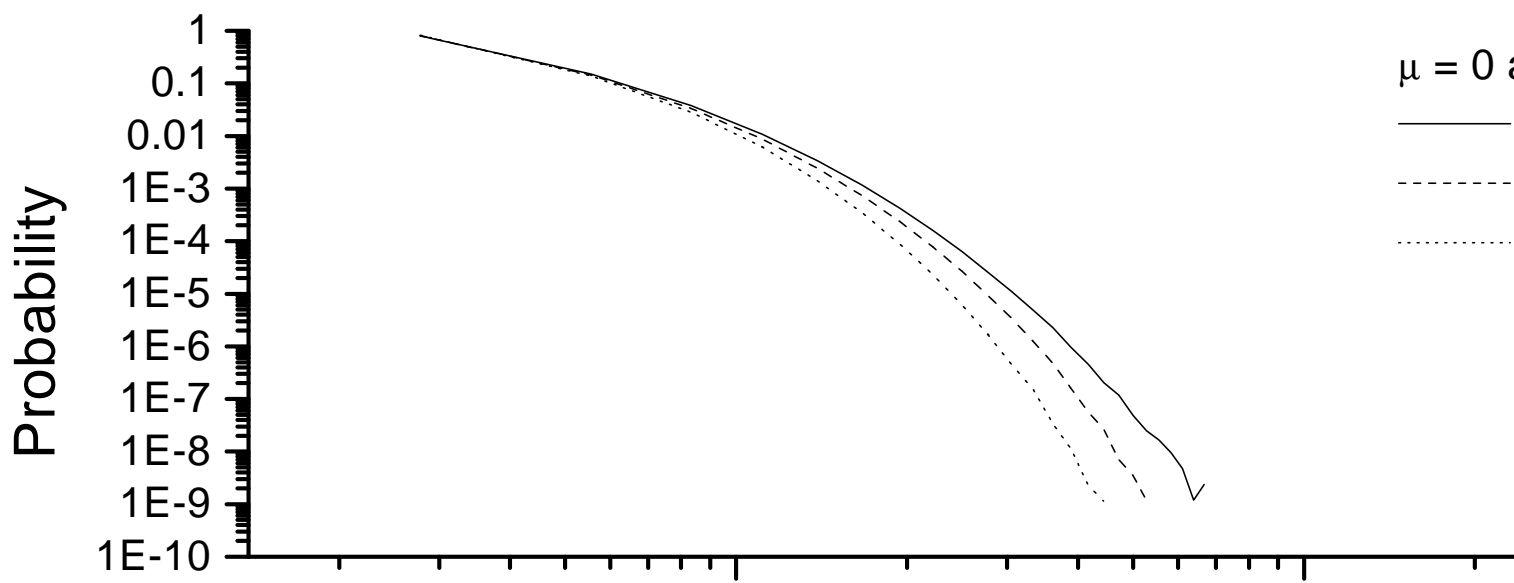
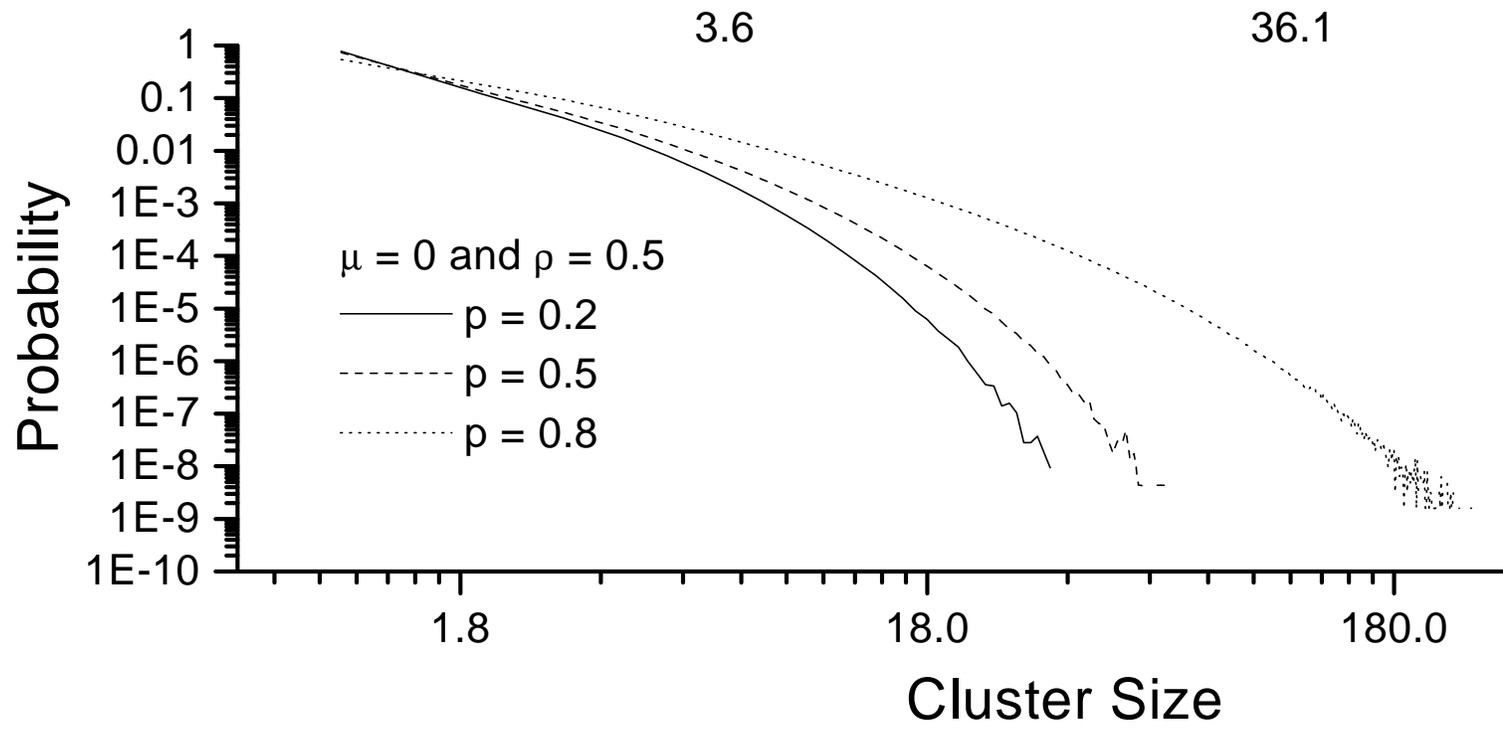
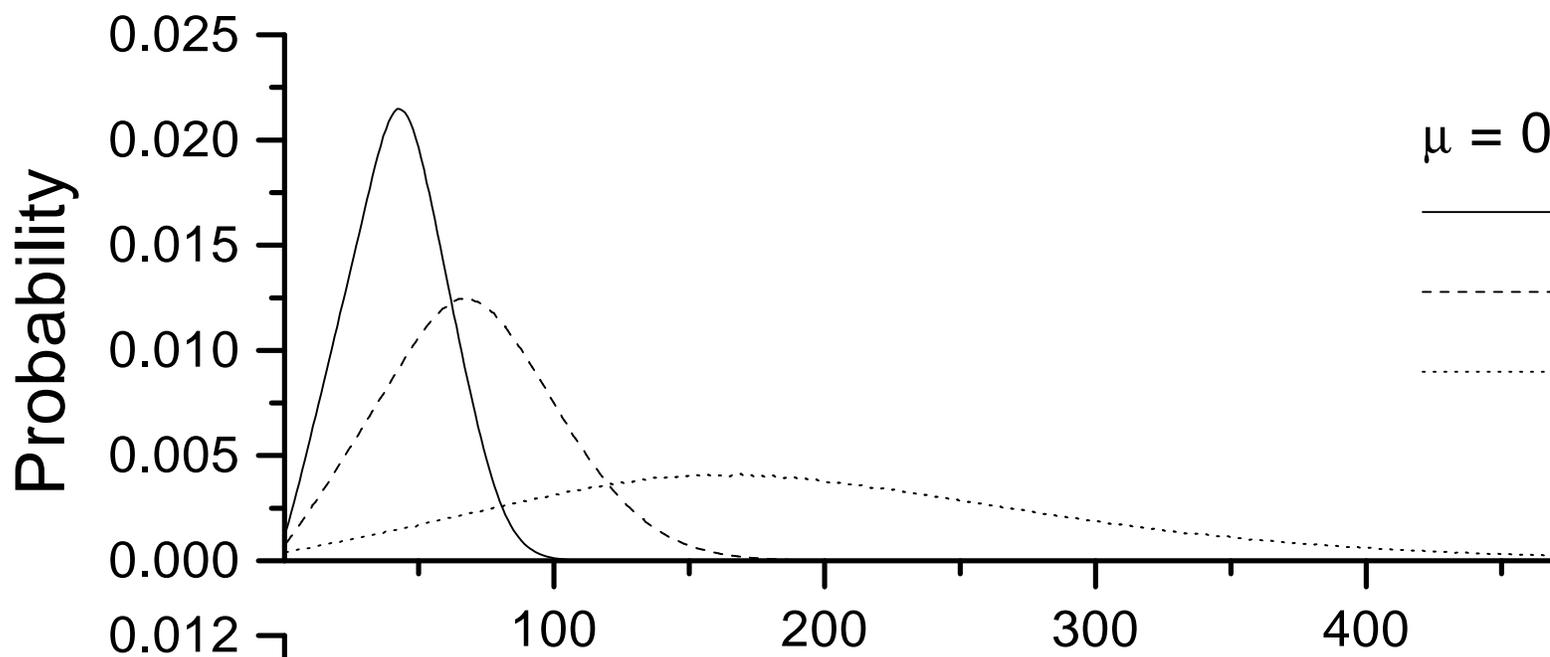
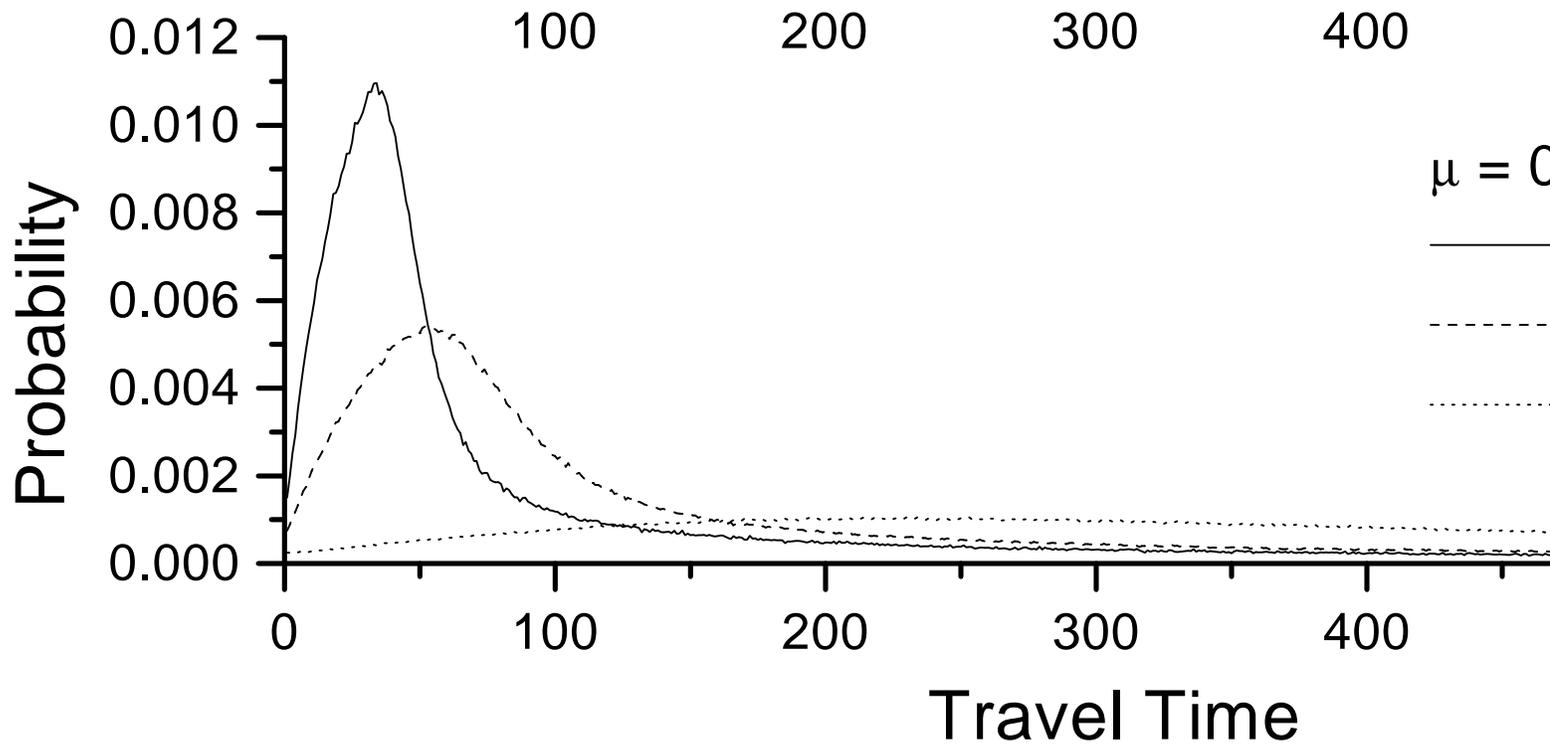
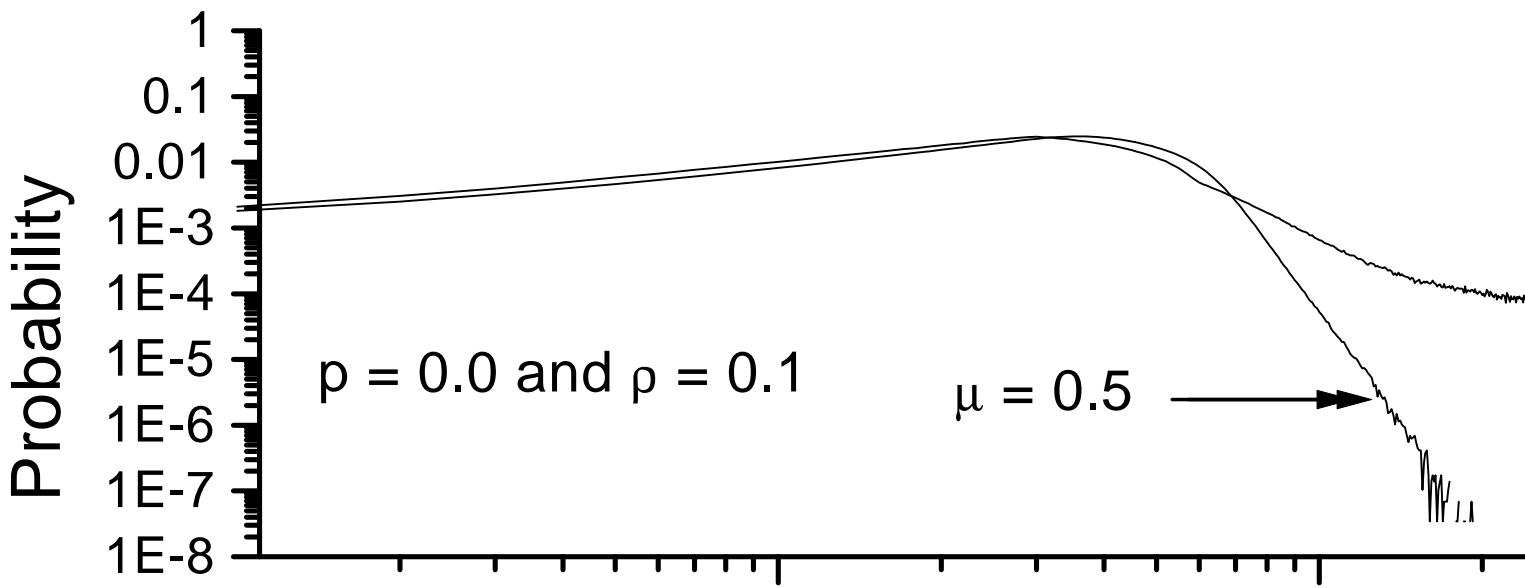
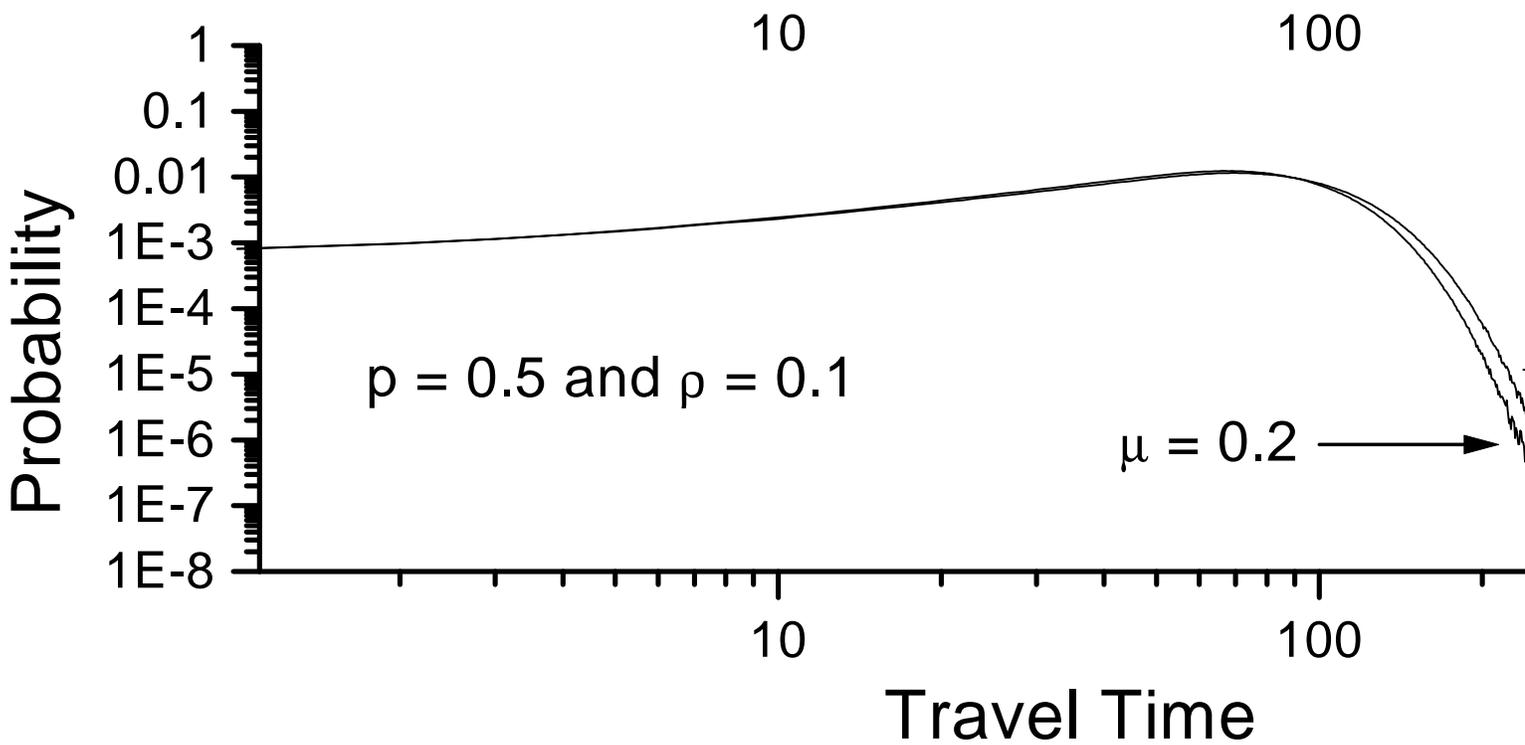
\end{document}